\documentclass{optica-article}

\journal{opticajournal} 

\articletype{Research Article}

\usepackage{pdfpages} 
\usepackage{pgffor} 

\makeatletter
\AtBeginDocument{\let\LS@rot\@undefined}
\makeatother

\def\supplementfilename{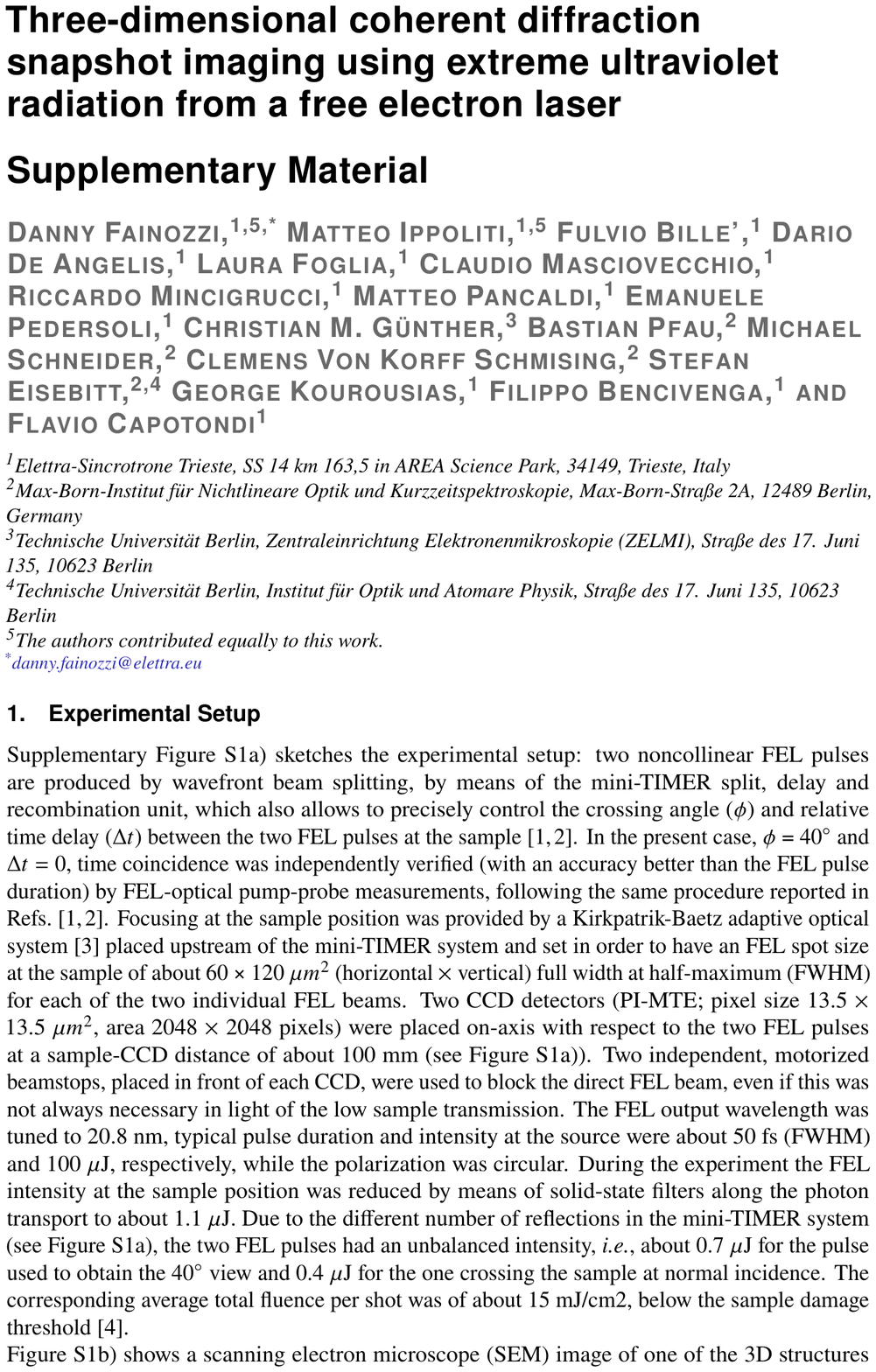}

\pdfximage{\supplementfilename}
\def\numbersupplementpages{\the\pdflastximagepages}

\newif\ifarXiv
\arXivtrue 

\begin{document}

\title{Three-dimensional coherent diffraction snapshot imaging using extreme ultraviolet radiation from a free electron laser}

\author{Danny Fainozzi,\authormark{1,5,*} Matteo Ippoliti,\authormark{1,5} Fulvio Bille’,\authormark{1} Dario De Angelis,\authormark{1} Laura Foglia,\authormark{1} Claudio Masciovecchio,\authormark{1} Riccardo Mincigrucci,\authormark{1} Matteo Pancaldi,\authormark{1} Emanuele Pedersoli,\authormark{1} Christian M. Günther,\authormark{3} Bastian Pfau,\authormark{2} Michael Schneider,\authormark{2} Clemens Von Korff Schmising,\authormark{2} Stefan Eisebitt,\authormark{2,4} George Kourousias,\authormark{1} Filippo Bencivenga,\authormark{1} and Flavio Capotondi\authormark{1}}

\address{\authormark{1}Elettra-Sincrotrone Trieste, SS 14 km 163,5 in AREA Science Park, 34149, Trieste, Italy\\
\authormark{2}Max-Born-Institut für Nichtlineare Optik und Kurzzeitspektroskopie, Max-Born-Straße 2A, 12489 Berlin, Germany\\
\authormark{3}Technische Universität Berlin, Zentraleinrichtung Elektronenmikroskopie (ZELMI), Straße des 17. Juni 135, 10623 Berlin\\
\authormark{4}Technische Universität Berlin, Institut für Optik und Atomare Physik, Straße des 17. Juni 135, 10623 Berlin\\
\authormark{5}The authors contributed equally to this work.}

\email{\authormark{*}danny.fainozzi@elettra.eu} 


\begin{abstract*} 
The possibility to obtain a three-dimensional representation of a single object with sub-µm resolution is crucial in many fields, from material science to clinical diagnostics. This is typically achieved through tomography, which combines multiple two-dimensional images of the same object captured at different orientations. However, this serial imaging method prevents single-shot acquisition in imaging experiments at free electron lasers. In the present experiment, we report on a new approach to 3D imaging using extreme-ultraviolet radiation. In this method, two EUV pulses hit simultaneously an isolated 3D object from different sides, generating independent coherent diffraction patterns, resulting in two distinct bidimensional views obtained via phase retrieval. These views are then used to obtain a 3D reconstruction using a ray tracing algorithm. This EUV stereoscopic imaging approach, similar to the natural process of binocular vision, provides sub-µm spatial resolution and single shot capability. Moreover, ultrafast time resolution and spectroscopy can be readily implemented, a further extension to X-ray wavelengths can be envisioned as well.

\end{abstract*}

\section{Introduction}
The three-dimensional (3D) structure of materials is often related to their properties and functionalities. The understanding of such relations at sub-$\mu$m length-scales is crucial for many applications, \textit{e.g.}, to control the tribological properties of micro-electro-mechanical systems \cite{rymuza1999control} or to study the electrochemical processes of corrosion and percolation that limit the lifetime of rechargeable batteries \cite{cui2021micronanostructured,shapiro2014chemical}. Short-wavelength probes are needed to achieve such high resolution. Among them, extreme-ultraviolet (EUV) and X-ray photons can be tuned to specific absorption edges, adding chemical selectivity and sensibility to the local environment \cite{jiang2013three,tripathi2011dichroic,kordel2020laboratory}, and have been used to obtain sub-µm resolution images of, \textit{e.g.}: non-periodic magnetic domains \cite{buttner2013magnetic,von2014imaging}, artificial nanostructures \cite{robinson2001reconstruction,gao2019three} and biological specimens \cite{nishino2009three,shapiro2005biological}.  Furthermore, the high brightness of EUV/X-ray free electron laser (FEL) sources has enabled the single-shot approach \cite{neutze2000potential,chapman2006femtosecond}, which is emerging as a crucial tool for probing samples that cannot withstand high radiation doses \cite{marchesini2003coherent,bielecki2020perspectives,sun2018current} and for studying irreversible dynamical processes \cite{barty2008ultrafast,seibert2011single}. Tomographic approaches based on a serial replacement of nearly identical targets \cite{loh2010cryptotomography,ekeberg2015three} and complex post-processing \cite{kommera2021accelerate} permit FEL-based 3D imaging. Alternatively, Geilhufe et al. \cite{geilhufe2014extracting} proposed to combine single frame information of Fourier-transform X-ray holography with numerical calculations of wave-front propagation to adjust the focus at different depths, in order to retrieve valuable 3D images. However, the combination of single-shot and 3D imaging was not yet reported, though potentially viable at FELs, as shown in the present manuscript.
The strategy followed here to obtain 3D information in a single-shot is to simultaneously determine multiple two-dimensional (2D) views of the same object, with time-coincident sample illumination from different view angles \cite{chang20213d,goldberger2020three}. A scheme for the simultaneous collection of X-ray images was theoretically proposed by Schmidt et al \cite{schmidt2008tomographic}, while the first experimental results in this direction have been reported by Duarte et al. \cite{duarte2019computed} using an EUV table-top source. In the present study a 3D visualization algorithm calculates disparity maps from two distinct EUV coherent diffraction imaging (CDI) views, producing a 3D point cloud reconstruction of the object. This approach requires small angles between the two views, in order to obtain accurate disparity maps. This limits the applicability to complex objects with obscured areas. Moreover, the reconstruction algorithm is merely based on geometrical projections, thus it does not convey any physical properties of the sample under investigation. Here we extend the work of Duarte et al. \cite{duarte2019computed} to demonstrate single-shot 3D imaging at FEL sources. In particular, we used a special split-recombination setup to illuminate the sample from different orientations and developed a reconstruction algorithm that incorporates physical information about the sample, rather than only relying on geometrical projections. This enables a larger separation angle between the two views, resulting in a more accurate determination of the 3D structure. The robustness of the algorithm allows us to reconstruct different 3D structures even if the object is exposed to a single FEL shot. This result paves the way to time-domain 3D imaging studies of, \textit{e.g.}, structural changes occurring in chemical reactions or biological processes, phase transitions or spin texture dynamics.

\section{Experiment}
\begin{figure}[ht!]
\centering\includegraphics[width=12cm]{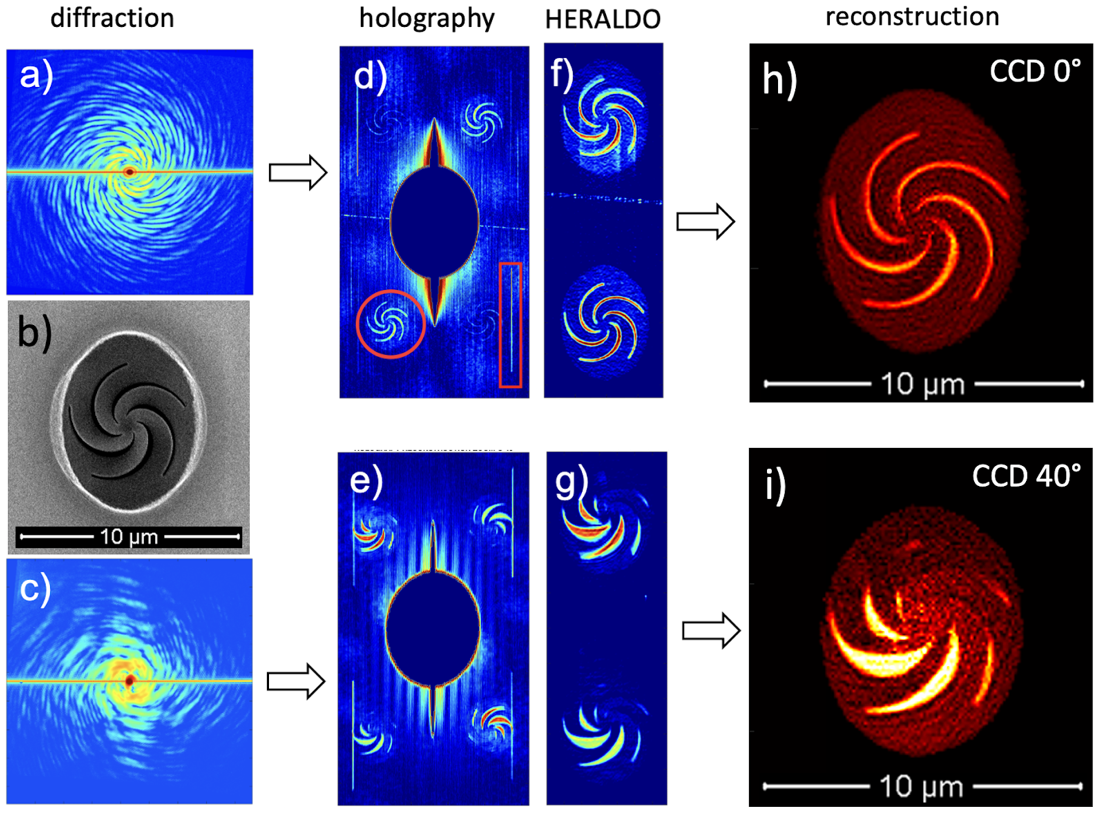}
\caption{Panel a) and c) show the diffraction patterns from the sample in panel  b) collected, respectively, at 0° and 40° view angle; the image in panel b) was recorded by scanning electron microscopy (SEM). Panels d) and e) display the holographic reconstruction of the sample (inside the red circle) and, at the same time, a glimpse on the shape of the support (red circle plus red rectangle). Panels f) and g) show the HERALDO reconstruction used as the first guess in the iterative phase retrieval algorithm, which led to the CDI reconstructed images in panels h) and i).}
\label{fig:fig1}
\end{figure}

The experiment has been carried out at the DiProI beamline located at the FERMI FEL facility (Trieste, Italy) \cite{capotondi2013invited,allaria2012highly}, using a system previously employed for FEL-based transient grating experiments and thoroughly described elsewhere \cite{bencivenga2015four,mincigrucci2018advances}. This setup  permitted  to generate two FEL pulses and recombine them at the sample, precisely setting the crossing angle ($\theta$ = 40$^{\circ}$) and ensuring time-coincidence. The spot size at the sample for each crossed FEL beam was about 60 × 120 $\mu m^2$ full width at half-maximum (FWHM), the FEL wavelength was 20.8 nm, the pulse duration 50 fs (FWHM) and the typical intensity at the sample of about  1.1 $\mu$J per shot, while the polarization was circular. We used as a sample different kinds of 3D structures, like the five-fold helicoidal structure shown in Fig. \ref{fig:fig1}b). The sample was oriented in order to have the surface orthogonal to one of the two FEL beams. Close to each sample there were properly-oriented holographic pin-holes and an extended HERALDO reference \cite{guizar2008direct,guizar2007holography}. Two CCD detectors were placed on-axis with respect to either of the two FEL beams at a distance of 100 mm from the sample. Hereafter we will refer to these two CCDs as CCD 40$^{\circ}$ and CCD 0$^{\circ}$, where the latter is the one placed on-axis with the FEL beam orthogonal to the sample surface. Further details on the setup and on the composition of the samples can be found in the Supplementary Material (SM).

\section{Results}
\begin{figure}[ht!]
\centering\includegraphics[width=12cm]{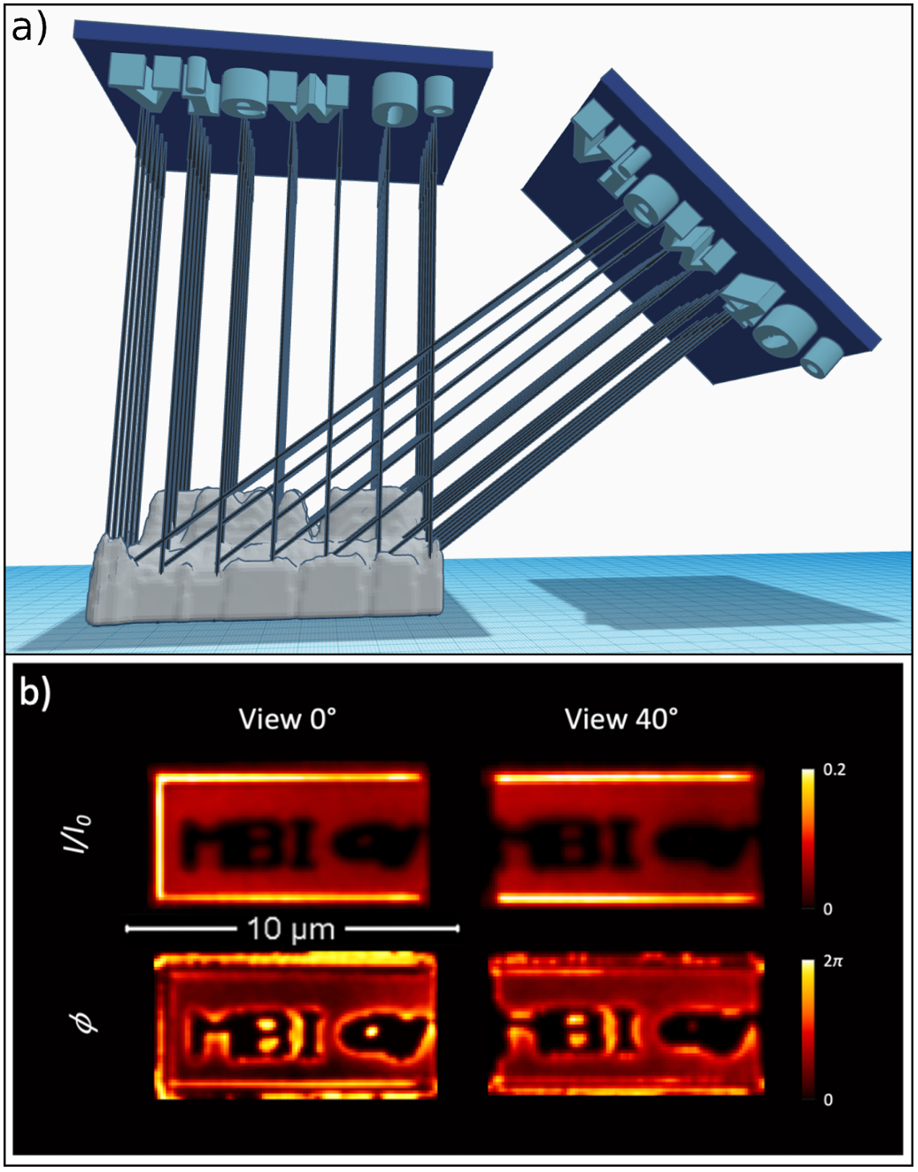}
\caption{Panel a) is a 3D rendering of the ray tracing reconstruction algorithm. Panel b) shows the $I/I_0$ and $\phi$ maps of the two CDI views, as computed from the 3D reconstructed sample.}
\label{fig:fig2}
\end{figure}

Figs. \ref{fig:fig1}a) and \ref{fig:fig1}c) display an example of diffraction patterns after post processing corrections (see section 2 of the SM) collected, respectively, by CCD 0$^{\circ}$ and CCD 40$^{\circ}$ from the sample shown in Figure \ref{fig:fig1}b). For each pair of such simultaneously collected diffraction patterns, we extract two holographic projections using the reference pinhole (Figs. \ref{fig:fig1}d) and \ref{fig:fig1}e)) and the HERALDO line (Figs. \ref{fig:fig1}f) and \ref{fig:fig1}g)). These holographic views are further refined using CDI reconstruction algorithms \cite{fienup1982phase,capotondi2012scheme}. In such a process, we use the holograms as input for the phase retrieval code and to define the support, which sets the spatial position of the unknown object during the CDI reconstruction of the diffracted phase. The 2D images obtained via the CDI reconstruction are presented in Fig. \ref{fig:fig1}h) and \ref{fig:fig1}i), for sample surface normal and tilted views respectively. In order to determine an absolute value of the sample transmission, during the CDI reconstruction, for each iteration the 2D images are renormalized by the average value of the clearance aperture in the sample (brighter regions in Figs \ref{fig:fig1}h) and \ref{fig:fig1}i)), \textit{i.e.}, the area where photons are not absorbed by the sample. The normal incidence view clearly resolves the five-fold geometry of the helicoidal structure, while the tilted view shows almost fully open structures, where the impinging photons does not interact with the sample (bottom left part of the image) and obscured areas (top right part of the image); here the lamellae forming the object overlap with the below substrate along the view direction due to the large view angle.
The 3D reconstruction algorithm uses the information contained in the two 2D CDI views, as illustrated in Fig. \ref{fig:fig2}a). These information are essentially the EUV transmission ($I/I_0$) and phase ($\phi$), as shown in Fig. \ref{fig:fig2}b) for the sample displayed in SM Fig. S1b). In the reconstruction process, the values $I/I_0$ and $\phi$ of the set of voxels along the directions defined by the two views must match the experimental data. The final solution is selected based on this matching, as evaluated by comparing both $I/I_0$ and $\phi$ of the experimental views with those computed from the 3D reconstructed sample. This evaluation is performed using a ray tracing approach, that assumes an aprioristic knowledge of sample thickness and composition, recently developed for the 3D stereographic reconstruction starting from 2D X-ray fluorescence images \cite{bille2016x,kourousias2019xrf,ippoliti2022reconstruction}. More information on the 2D and 3D reconstruction algorithms, as well as on their relation, can be found in the SM.

\begin{figure}[ht!]
\centering\includegraphics[width=12cm]{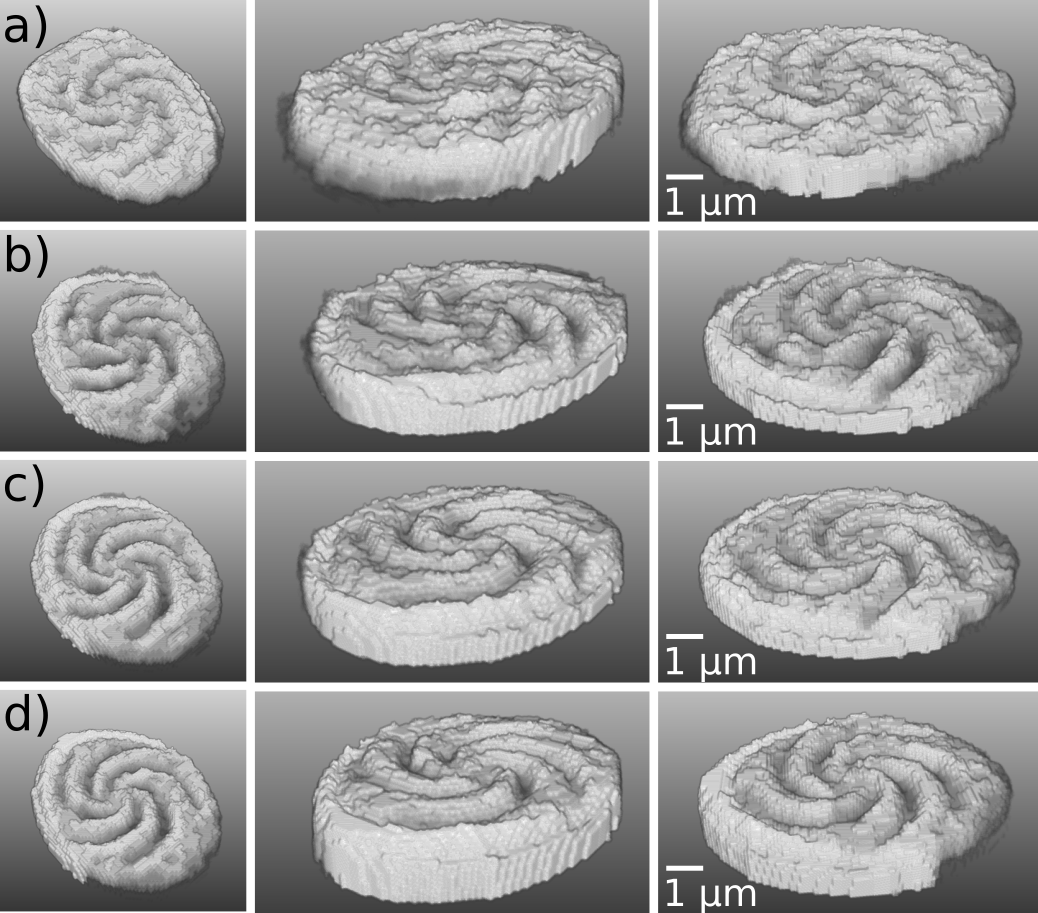}
\caption{3D reconstructions of the sample shown in Fig. \ref{fig:fig1}b), as obtained from a single-shot exposure (panel a)), by averaging two different single shot exposures (panel b)), from 25 shots exposures (panel c)) and by averaging 30 different dataset each of them having 25 shots exposures (panel d)).}
\label{fig:fig3}
\end{figure}

\noindent
Fig. \ref{fig:fig3} shows an example of 3D reconstruction of the five-fold helicoidal structure shown in Fig. \ref{fig:fig1}b). These reconstructions have been obtained using different datasets collected by exposing the sample to a different number of FEL shots, from one single-shot up to 750 shots. Other 3D reconstructed objects, including the sample displayed in Fig. \ref{fig:fig2}b), are shown in the SM. In all cases the combination of holographically-guided CDI and our 3D ray tracing projection algorithms is able to retrieve a realistic stereographic solution of the sample. As quantitatively discussed further below, the image quality does not substantially degrade on reducing the number of exposures, highlighting the computational robustness of the stereographic retrieval procedure.

\section{Discussion}
\begin{figure}[ht!]
\centering\includegraphics[width=14cm]{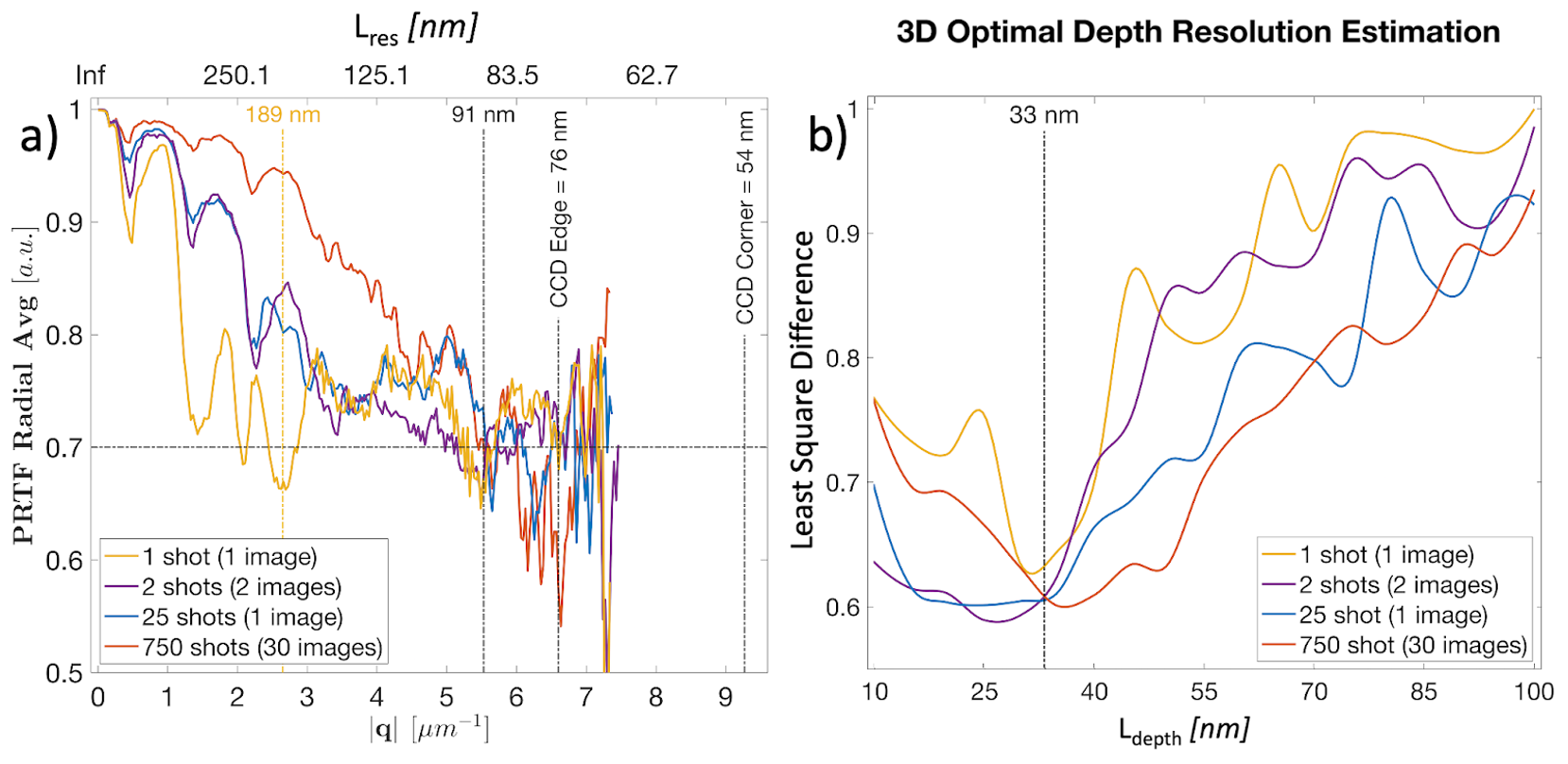}
\caption{Panel a) shows the radially averaged PRTF for the reconstructions displayed in Fig. \ref{fig:fig3}. Yellow, violet, blue and red lines refer to reconstructions based on, respectively, a single-shot exposure, 2 averaged single shot exposures, a single 25 shot exposure and the average of 30 different 25 shots exposures. Panel b) shows the normalized sum of least squares as a function of $\Delta z$ for the same reconstructions considered in panel a). The vertical line indicates the estimated optimal resolution along the z coordinate (sample depth).}
\label{fig:fig4}
\end{figure}

In order to provide a quantitative estimate of the lateral resolution ($L_{\text{res}}$), we computed the phase retrieval transfer function (PRTF) in the 1060 iterations of the CDI code. 
Fig. \ref{fig:fig4}a) shows the comparison of the PRFT for the five-fold helicoidal structure displayed in Fig. \ref{fig:fig1}b). In this analysis we consider reconstructions under the different exposure conditions shown in Fig. \ref{fig:fig3}, namely: (i) single-shot, (ii) averaging two images each one corresponding to a single-shot acquisition, (iii) a 25-shot acquisition and (iv) averaging 30 images each one corresponding to a 25 shots acquisition. The value of the PRFT is well above the 1/e threshold suggested by Chapman et al. \cite{chapman2006femtosecond} for all the range in the exchanged momentum |q| recorded by the detector (details on the |q|-range are in the SM). We ascribe this to the more uniform coverage of the detector plane by the diffraction pattern originating by our circularly symmetric object compared to the lower-symmetry one considered by Chapman, resulting in a larger number of pixels for which the phase can be retrieved accurately. Therefore, we set the threshold for defining the resolution at PRFT = 0.7, roughly matching the high-|q| asymptotic value of the curves reported in Figs. 4a), which is likely to be limited by the SNR in such a high-|q| range. Different samples show different asymptotic values, as discussed in the SM. The PRTF function is larger than 0.7 in almost the entire |q|-range, for reconstructions obtained from datasets corresponding to a different number of shots. This indicates that the actual resolution does not substantially decrease from the multi-shot to the single-shot regime. According to the aforementioned empiric PRTF = 0.7 level, the lateral resolution ranges from about 90 $\pm$ 10 nm for multi-shot averaged images to about 190 $\pm$ 20 nm for the single-shot.

\noindent
The depth resolution ($L_{\text{depth}}$) along the z-coordinate (sample depth) for the 3D reconstructed sample shown in Fig. \ref{fig:fig3} was evaluated by varying the sampling step along z, from $\Delta z$ = 10 nm to 100 nm, and generating an ensemble of associated solutions. These are then evaluated through the structural similarity index measure (SSIM) \cite{wang2004image}, as described in section 5 of the SM. The SSIM analysis isolates the solutions that give the most similar output to the input 2D views, in terms of the least squares difference between the simulated and the reconstructed $I/I_0$ and $\phi$ maps. We defined the resolution in the z-axis as the value of $\Delta z$ in correspondence of which the least squares differences reach a minimum, meaning that the two CDI reconstructed views (both in amplitude and in phase) do not provide additional information to the ray tracing stereographic code. Fig. \ref{fig:fig4}b) shows the dependence on $\Delta z$ of the least squares difference of the likelihood reconstruction selected by SSIM analysis for the four 3D reconstructions reported in Fig. \ref{fig:fig3}, which are associated to different numbers of FEL shots. The SSIM analysis shows a similar value of $L_{\text{depth}}$ = 33 $\pm$ 5 nm, nearly independent from the number of shots. This suggests that 3D ray tracing algorithms based on the aprioristic knowledge of material properties (absorption length, layer thickness) are less sensitive to the natural degradation of the signal to noise ratio due to the reduced number of exposures. This observation is consistent with recent numerical simulations, where the 3D reconstruction process recovers the correct topography even if the input thickness maps have spatially homogeneous white noise \cite{ippoliti2022reconstruction}.

\section{Conclusions}
We combined a non-collinear EUV split-recombination setup (compact, conceptually simple and easily reproducible) with a data processing protocol based on both CDI phase retrieval and ray tracing algorithms, to experimentally demonstrate the capability to determine the 3D structure of an object from a single-flash of EUV light (20.8 nm wavelength) at moderate fluence, with a resolution in the order of 100 nm. This allowed us to perform the experiment well below the typical damage threshold of most materials. Though the structures used in this pilot experiment have been realized ad hoc, they are quite representative of possible samples, e.g., magnetic nanostructures or biological samples, in terms of size of structural details (10’s-100’s nm) and overall dimensions (a few µm), indicating how this approach can be readily used to investigate such a class of samples. This approach can be profitably employed in time-resolved experiments, where high efficiency is crucial, and, in case of non-reproducible dynamics, inherently necessary. The introduction of an optical pulse (for pump-probe experiments) in this setup is straightforward, since it is routinely employed in EUV transient grating experiments \cite{bencivenga2015four,mincigrucci2018advances}. Such experiments were performed up to about 100 eV photon energy and showed how the system is compatible with spectroscopic approaches, where the EUV wavelength is scanned across specific resonances of the material \cite{bohinc2019nonlinear}, thus enabling element-selective 3D imaging. Extension to higher photon energies, up to the water window (300–500 eV) and beyond, towards the L-edges of 3d transition metals (600–1000 eV), is technically possible by exploiting state-of-the-art multilayer coatings \cite{yulin2021coatings}. Advances in the fabrication of transmissive X-ray diffraction optics \cite{kujala2019characterizing,rosner2018exploiting} enable us to envision split-recombination systems based on such optics, making this approach compatible even with hard X-rays. X-ray diffractive optics may allow handling several replicas of FEL pulses simultaneously impinging on the sample from more than two different angles \cite{chang20213d,goldberger2020three,kharitonov2022single}. This multiplex approach can greatly improve 3D reconstruction, that in the present version assumes that the sample composition (constituent materials and density) is known a priori, and the absence of gaps along the photon propagation direction. These assumptions can be reasonably met in many practical cases. However, multiple views could overcome the need for these assumptions to be met. For instance, for a pure absorbing sample, if the composition is uncertain it would be hard to determine from a limited number of views whether the light is traversing a shorter path of high absorbing material or a longer path through a lower absorbing one. Since both phase and amplitude are derived in CDI reconstruction the complex scattering quotient, \textit{i.e.}, the ratio of reconstructed phase and the natural logarithm of the reconstructed amplitude, can be used to determine the local material content \cite{eschen2022material}. This uncertainty in the sample composition could also account for the small differences observed between the real 3D samples used in the present work and the reconstructed models, since a given density and composition was assumed. The ray tracing approach presented here is a promising way to reconstruct 3D samples as it has already shown robustness and flexibility with the minimal (two) viewing angles involved in this experiment and enables large angular separation between views. The system can be readily exploited in time-domain studies, extended to include EUV spectroscopy and can surely be improved by exploiting more views, either by using multiple time and space coincident beams or, if single shot is not required, by rotating the sample.

\section{Backmatter}

\begin{backmatter}
\bmsection{Funding}
Content in the funding section will be generated entirely from details submitted to Prism. Authors may add placeholder text in the manuscript to assess length, but any text added to this section in the manuscript will be replaced during production and will display official funder names along with any grant numbers provided. If additional details about a funder are required, they may be added to the Acknowledgments, even if this duplicates information in the funding section. See the example below in Acknowledgements. For preprint submissions, please include funder names and grant numbers in the manuscript.

\bmsection{Acknowledgments}
Authors acknowledge the FERMI team for invaluable support and Hamed Merdji and his team for the fruitful discussions on stereo imaging based on disparity map.

\bmsection{Disclosures}
The authors declare no conflicts of interest.

\bmsection{Data availability} Data underlying the results presented in this paper are not publicly available at this time but may be obtained from the authors upon reasonable request.

\bmsection{Supplemental document}
See Supplement 1 for supporting content. 

\end{backmatter}




\bibliography{bib}

\begin{thebibliography}{10}
\newcommand{\enquote}[1]{``#1''}

\bibitem{rymuza1999control}
Z.~Rymuza, \enquote{Control tribological and mechanical properties of mems
  surfaces. part 1: critical review,} {\protect\JournalTitle{Microsystem
  Technologies}} \textbf{5}, 173--180 (1999).

\bibitem{cui2021micronanostructured}
B.-F. Cui, X.-P. Han, and W.-B. Hu, \enquote{Micronanostructured design of
  dendrite-free zinc anodes and their applications in aqueous zinc-based
  rechargeable batteries,} {\protect\JournalTitle{Small Structures}}
  \textbf{2}, 2000128 (2021).

\bibitem{shapiro2014chemical}
D.~A. Shapiro, Y.-S. Yu, T.~Tyliszczak, J.~Cabana, R.~Celestre, W.~Chao,
  K.~Kaznatcheev, A.~D. Kilcoyne, F.~Maia, S.~Marchesini, Y.~S. Meng,
  T.~Warwick, L.~L. Yang, and H.~A. Padmore, \enquote{Chemical composition
  mapping with nanometre resolution by soft x-ray microscopy,}
  {\protect\JournalTitle{Nature Photonics}} \textbf{8}, 765--769 (2014).

\bibitem{jiang2013three}
H.~Jiang, R.~Xu, C.-C. Chen, W.~Yang, J.~Fan, X.~Tao, C.~Song, Y.~Kohmura,
  T.~Xiao, Y.~Wang, Y.~Fei, T.~Ishikawa, W.~L. Mao, and J.~Miao,
  \enquote{Three-dimensional coherent x-ray diffraction imaging of molten iron
  in mantle olivine at nanoscale resolution,} {\protect\JournalTitle{Physical
  Review Letters}} \textbf{110}, 205501 (2013).

\bibitem{tripathi2011dichroic}
A.~Tripathi, J.~Mohanty, S.~H. Dietze, O.~G. Shpyrko, E.~Shipton, E.~E.
  Fullerton, S.~S. Kim, and I.~McNulty, \enquote{Dichroic coherent diffractive
  imaging,} {\protect\JournalTitle{Proceedings of the National Academy of
  Sciences}} \textbf{108}, 13393--13398 (2011).

\bibitem{kordel2020laboratory}
M.~K{\"o}rdel, A.~Dehlinger, C.~Seim, U.~Vogt, E.~Fogelqvist, J.~A. Sellberg,
  H.~Stiel, and H.~M. Hertz, \enquote{Laboratory water-window x-ray
  microscopy,} {\protect\JournalTitle{Optica}} \textbf{7}, 658--674 (2020).

\bibitem{buttner2013magnetic}
F.~B{\"u}ttner, C.~Moutafis, A.~Bisig, P.~Wohlh{\"u}ter, C.~M. G{\"u}nther,
  J.~Mohanty, J.~Geilhufe, M.~Schneider, C.~v.~K. Schmising, S.~Schaffert,
  B.~Pfau, M.~Hantschmann, M.~Riemeier, M.~Emmel, S.~Finizio, G.~Jakob,
  M.~Weigand, J.~Rhensius, j.~H. Franken, R.~Lavrijsen, H.~J.~M. Swagten,
  H.~Stoll, S.~Eisebitt, and M.~Kläui, \enquote{Magnetic states in low-pinning
  high-anisotropy material nanostructures suitable for dynamic imaging,}
  {\protect\JournalTitle{Physical Review B}} \textbf{87}, 134422 (2013).

\bibitem{von2014imaging}
C.~von Korff~Schmising, B.~Pfau, M.~Schneider, C.~G{\"u}nther, M.~Giovannella,
  J.~Perron, B.~Vodungbo, L.~M{\"u}ller, F.~Capotondi, E.~Pedersoli, N.~Mahne,
  J.~Lüning, and S.~Eisebitt, \enquote{Imaging ultrafast demagnetization
  dynamics after a spatially localized optical excitation,}
  {\protect\JournalTitle{Physical review letters}} \textbf{112}, 217203 (2014).

\bibitem{robinson2001reconstruction}
I.~K. Robinson, I.~A. Vartanyants, G.~Williams, M.~Pfeifer, and J.~Pitney,
  \enquote{Reconstruction of the shapes of gold nanocrystals using coherent
  x-ray diffraction,} {\protect\JournalTitle{Physical review letters}}
  \textbf{87}, 195505 (2001).

\bibitem{gao2019three}
Y.~Gao, R.~Harder, S.~H. Southworth, J.~R. Guest, X.~Huang, Z.~Yan, L.~E.
  Ocola, Y.~Yifat, N.~Sule, P.~J. Ho, M.~Pelton, N.~F. Scherer, and L.~Young,
  \enquote{Three-dimensional optical trapping and orientation of microparticles
  for coherent x-ray diffraction imaging,} {\protect\JournalTitle{Proceedings
  of the National Academy of Sciences}} \textbf{116}, 4018--4024 (2019).

\bibitem{nishino2009three}
Y.~Nishino, Y.~Takahashi, N.~Imamoto, T.~Ishikawa, and K.~Maeshima,
  \enquote{Three-dimensional visualization of a human chromosome using coherent
  x-ray diffraction,} {\protect\JournalTitle{Physical review letters}}
  \textbf{102}, 018101 (2009).

\bibitem{shapiro2005biological}
D.~Shapiro, P.~Thibault, T.~Beetz, V.~Elser, M.~Howells, C.~Jacobsen, J.~Kirz,
  E.~Lima, H.~Miao, A.~M. Neiman, and D.~Sayre, \enquote{Biological imaging by
  soft x-ray diffraction microscopy,} {\protect\JournalTitle{Proceedings of the
  National Academy of Sciences}} \textbf{102}, 15343--15346 (2005).

\bibitem{neutze2000potential}
R.~Neutze, R.~Wouts, D.~Van Der~Spoel, E.~Weckert, and J.~Hajdu,
  \enquote{Potential for biomolecular imaging with femtosecond x-ray pulses,}
  {\protect\JournalTitle{Nature}} \textbf{406}, 752--757 (2000).

\bibitem{chapman2006femtosecond}
H.~N. Chapman, A.~Barty, M.~J. Bogan, S.~Boutet, M.~Frank, S.~P. Hau-Riege,
  S.~Marchesini, B.~W. Woods, S.~Bajt, W.~H. Benner, R.~A. London, E.~Plönjes,
  M.~Kuhlmann, R.~Treusch, S.~Düsterer, T.~Tschentscher, J.~R. Schneider,
  E.~Spiller, T.~Möller, C.~Bostedt, T.~Hoener, D.~A. Shapiro, K.~O. Hodgson,
  D.~Van Der~Spoel, F.~Burmeister, M.~Bergh, C.~Caleman, G.~Huldt, M.~M.
  Seibert, F.~R. N.~C. Maia, R.~W. Lee, A.~Szöke, N.~Timneanu, and J.~Hajdu,
  \enquote{Femtosecond diffractive imaging with a soft-x-ray free-electron
  laser,} {\protect\JournalTitle{Nature Physics}} \textbf{2}, 839--843 (2006).

\bibitem{marchesini2003coherent}
S.~Marchesini, H.~Chapman, S.~Hau-Riege, R.~London, A.~Szoke, H.~He,
  M.~Howells, H.~Padmore, R.~Rosen, J.~Spence, and U.~Weierstall,
  \enquote{Coherent x-ray diffractive imaging: applications and limitations,}
  {\protect\JournalTitle{Optics Express}} \textbf{11}, 2344--2353 (2003).

\bibitem{bielecki2020perspectives}
J.~Bielecki, F.~R. Maia, and A.~P. Mancuso, \enquote{Perspectives on single
  particle imaging with x rays at the advent of high repetition rate x-ray free
  electron laser sources,} {\protect\JournalTitle{Structural Dynamics}}
  \textbf{7}, 040901 (2020).

\bibitem{sun2018current}
Z.~Sun, J.~Fan, H.~Li, and H.~Jiang, \enquote{Current status of single particle
  imaging with x-ray lasers,} {\protect\JournalTitle{Applied Sciences}}
  \textbf{8}, 132 (2018).

\bibitem{barty2008ultrafast}
A.~Barty, S.~Boutet, M.~J. Bogan, S.~Hau-Riege, S.~Marchesini,
  K.~Sokolowski-Tinten, N.~Stojanovic, R.~Tobey, H.~Ehrke, A.~Cavalleri,
  S.~Düsterer, M.~Frank, S.~Bajt, B.~W. Woods, M.~M. Seibert, J.~Hajdu,
  R.~Treusch, and H.~N. Chapman, \enquote{Ultrafast single-shot diffraction
  imaging of nanoscale dynamics,} {\protect\JournalTitle{Nature photonics}}
  \textbf{2}, 415--419 (2008).

\bibitem{seibert2011single}
M.~M. Seibert, T.~Ekeberg, F.~R. Maia, M.~Svenda, J.~Andreasson,
  O.~J{\"o}nsson, D.~Odi{\'c}, B.~Iwan, A.~Rocker, D.~Westphal, M.~Hantke,
  D.~P. DePonte, A.~Barty, J.~Schulz, L.~Gumprecht, N.~Coppola, A.~Aquila,
  M.~Liang, T.~A. White, A.~Martin, C.~Caleman, S.~Stern, C.~Abergel,
  V.~Seltzer, J.~M. Claverie, C.~Bostedt, J.~D. Bozek, D.~Boutet, A.~A.
  Miahnahri, M.~Messerschmidt, J.~Krzywinski, G.~Williams, K.~O. Hodgson, M.~J.
  Bogan, C.~Y. Hampton, R.~G. Sierra, D.~Starodub, I.~Andersson, S.~Bajt,
  M.~Barthelmess, J.~C.~H. Spence, P.~Fromme, U.~Weierstall, R.~Kirian,
  M.~Hunter, R.~B. Doak, R.~Marchesini, S.~P. Hau-Riege, M.~Frank, R.~L.
  Shoeman, L.~Lomb, S.~W. Epp, R.~Hartmann, D.~Rolles, A.~Rudenko, C.~Schmidt,
  L.~Foucar, N.~Kimmel, P.~Holl, B.~Rudek, B.~Erk, A.~Hömke, A.~Reich,
  D.~Pietschner, G.~Weidenspointner, L.~Strüder, G.~Hauser, H.~Gorke,
  J.~Ullrich, I.~Schlichting, S.~Herrmann, G.~Schaller, F.~Schopper, H.~Soltau,
  K.~U. Kühnel, R.~Andritschke, C.~D. Schröter, F.~Krasniqi, M.~Bott,
  S.~Schorb, D.~Rupp, M.~Adolph, T.~Gorkhover, H.~Hirsemann, G.~Potdevin,
  H.~Graafsma, B.~Nilsson, H.~N. Chapman, and J.~Hajdu, \enquote{Single
  mimivirus particles intercepted and imaged with an x-ray laser,}
  {\protect\JournalTitle{Nature}} \textbf{470}, 78--81 (2011).

\bibitem{loh2010cryptotomography}
N.~Loh, M.~J. Bogan, V.~Elser, A.~Barty, S.~Boutet, S.~Bajt, J.~Hajdu,
  T.~Ekeberg, F.~R. Maia, J.~Schulz, M.~M. Seibert, B.~Iwan, N.~Timneanu,
  S.~Marchesini, I.~Schlichting, R.~L. Shoeman, L.~Lomb, M.~Frank, M.~Liang,
  and H.~N. Chapman, \enquote{Cryptotomography: reconstructing 3d fourier
  intensities from randomly oriented single-shot diffraction patterns,}
  {\protect\JournalTitle{Physical review letters}} \textbf{104}, 225501 (2010).

\bibitem{ekeberg2015three}
T.~Ekeberg, M.~Svenda, C.~Abergel, F.~R. Maia, V.~Seltzer, J.-M. Claverie,
  M.~Hantke, O.~J{\"o}nsson, C.~Nettelblad, G.~Van Der~Schot, M.~Liang, D.~P.
  DePonte, A.~Barty, M.~M. Seibert, B.~Iwan, I.~Andersson, N.~D. Loh, A.~V.
  Martin, H.~N. Chapman, C.~Bostedt, J.~D. Bozek, k.~R. Ferguson, S.~W.
  Krzywinski, J~andEpp, D.~Rolles, A.~Rudenko, R.~K. Hartmann, N.~Kimmel, and
  J.~Hajdu, \enquote{Three-dimensional reconstruction of the giant mimivirus
  particle with an x-ray free-electron laser,} {\protect\JournalTitle{Physical
  review letters}} \textbf{114}, 098102 (2015).

\bibitem{kommera2021accelerate}
P.~R. Kommera, V.~B. Ramakrishnaiah, and C.~M. Sweeney, \enquote{Accelerate
  m-tip on gpus and deploy to summit and nersc-9 (against simulated data) wbs
  2.2. 4.05 exafel, milestone adse13-199,} Tech. rep., Los Alamos National
  Lab.(LANL), Los Alamos, NM (United States) (2021).

\bibitem{geilhufe2014extracting}
J.~Geilhufe, C.~Tieg, B.~Pfau, C.~G{\"u}nther, E.~Guehrs, S.~Schaffert, and
  S.~Eisebitt, \enquote{Extracting depth information of 3-dimensional
  structures from a single-view x-ray fourier-transform hologram,}
  {\protect\JournalTitle{Optics Express}} \textbf{22}, 24959--24969 (2014).

\bibitem{chang20213d}
C.~Chang, X.~Pan, H.~Tao, C.~Liu, S.~P. Veetil, and J.~Zhu, \enquote{3d
  single-shot ptychography with highly tilted illuminations,}
  {\protect\JournalTitle{Optics Express}} \textbf{29}, 30878--30891 (2021).

\bibitem{goldberger2020three}
D.~Goldberger, J.~Barolak, C.~G. Durfee, and D.~E. Adams,
  \enquote{Three-dimensional single-shot ptychography,}
  {\protect\JournalTitle{Optics Express}} \textbf{28}, 18887--18898 (2020).

\bibitem{schmidt2008tomographic}
K.~Schmidt, J.~Spence, U.~Weierstall, R.~Kirian, X.~Wang, D.~Starodub,
  H.~Chapman, M.~Howells, and R.~Doak, \enquote{Tomographic femtosecond x-ray
  diffractive imaging,} {\protect\JournalTitle{Physical review letters}}
  \textbf{101}, 115507 (2008).

\bibitem{duarte2019computed}
J.~Duarte, R.~Cassin, J.~Huijts, B.~Iwan, F.~Fortuna, L.~Delbecq, H.~Chapman,
  M.~Fajardo, M.~Kovacev, W.~Boutu, and H.~Merdji, \enquote{Computed stereo
  lensless x-ray imaging,} {\protect\JournalTitle{Nature Photonics}}
  \textbf{13}, 449--453 (2019).

\bibitem{capotondi2013invited}
F.~Capotondi, E.~Pedersoli, N.~Mahne, R.~Menk, G.~Passos, L.~Raimondi,
  C.~Svetina, G.~Sandrin, M.~Zangrando, M.~Kiskinova, S.~Bajt, M.~Barthelmess,
  H.~Fleckenstein, H.~N. Chapman, J.~Schulz, J.~Bach, R.~Frömter,
  S.~Schleitzer, L.~Müller, C.~Gutt, and G.~Grübel, \enquote{Invited article:
  Coherent imaging using seeded free-electron laser pulses with variable
  polarization: First results and research opportunities,}
  {\protect\JournalTitle{Review of scientific instruments}} \textbf{84}, 051301
  (2013).

\bibitem{allaria2012highly}
E.~Allaria, R.~Appio, L.~Badano, W.~Barletta, S.~Bassanese, S.~Biedron,
  A.~Borga, E.~Busetto, D.~Castronovo, P.~Cinquegrana, S.~Cleva, D.~Cocco,
  M.~Cornacchia, P.~Craievich, I.~Cudin, G.~D’Auria, M.~Dal~Forno, m.~B.
  Danailov, R.~De~Monte, G.~De~Ninno, P.~Delgiusto, A.~Demidovich, S.~Di~Mitri,
  B.~Diviacco, A.~Fabris, R.~Fabris, W.~Fawley, M.~Ferianis, E.~Ferrari,
  S.~Ferry, L.~Froehlich, P.~Furlan, G.~Gaio, F.~Gelmetti, L.~Giannessi,
  M.~Giannini, R.~Gobessi, R.~Ivanov, E.~Karantzoulis, M.~Lonza, A.~Lutman,
  B.~Mahieu, M.~Milloch, S.~V. Milton, M.~Musardo, I.~Nikolov, S.~Noe,
  G.~Parmigiani, F ang~Penco, M.~Petronio, L.~Pivetta, M.~Predonzani, F.~Rossi,
  L.~Rumiz, A.~Salom, C.~Scafuri, C.~Serpico, P.~Sigalotti, S.~Spampinati,
  C.~Spezzani, M.~Svandrlik, C.~Svetina, S.~Tazzari, M.~Trovo, R.~Umer,
  A.~Vascotto, M.~Veronese, R.~Visintini, M.~Zaccaria, D.~Zangrando, and
  M.~Zangrando, \enquote{Highly coherent and stable pulses from the fermi
  seeded free-electron laser in the extreme ultraviolet,}
  {\protect\JournalTitle{Nature Photonics}} \textbf{6}, 699--704 (2012).

\bibitem{bencivenga2015four}
F.~Bencivenga, R.~Cucini, F.~Capotondi, A.~Battistoni, R.~Mincigrucci,
  E.~Giangrisostomi, A.~Gessini, M.~Manfredda, I.~Nikolov, E.~Pedersoli,
  E.~Principi, C.~Svetina, P.~Parisse, F.~Casolari, M.~B. Danailov,
  M.~Kiskinova, and C.~Masciovecchio, \enquote{Four-wave mixing experiments
  with extreme ultraviolet transient gratings,} {\protect\JournalTitle{Nature}}
  \textbf{520}, 205--208 (2015).

\bibitem{mincigrucci2018advances}
R.~Mincigrucci, L.~Foglia, D.~Naumenko, E.~Pedersoli, A.~Simoncig, R.~Cucini,
  A.~Gessini, M.~Kiskinova, G.~Kurdi, N.~Mahne, M.~Manfredda, I.~P. Nikolov,
  E.~Principi, L.~Raimondi, M.~Zangrando, C.~Masciovecchio, F.~Capotondi, and
  F.~Bencivenga, \enquote{Advances in instrumentation for fel-based
  four-wave-mixing experiments,} {\protect\JournalTitle{Nuclear Instruments and
  Methods in Physics Research Section A: Accelerators, Spectrometers, Detectors
  and Associated Equipment}} \textbf{907}, 132--148 (2018).

\bibitem{guizar2008direct}
M.~Guizar-Sicairos and J.~R. Fienup, \enquote{Direct image reconstruction from
  a fourier intensity pattern using heraldo,} {\protect\JournalTitle{Optics
  letters}} \textbf{33}, 2668--2670 (2008).

\bibitem{guizar2007holography}
M.~Guizar-Sicairos and J.~R. Fienup, \enquote{Holography with extended
  reference by autocorrelation linear differential operation,}
  {\protect\JournalTitle{Optics express}} \textbf{15}, 17592--17612 (2007).

\bibitem{fienup1982phase}
J.~R. Fienup, \enquote{Phase retrieval algorithms: a comparison,}
  {\protect\JournalTitle{Applied optics}} \textbf{21}, 2758--2769 (1982).

\bibitem{capotondi2012scheme}
F.~Capotondi, E.~Pedersoli, M.~Kiskinova, A.~Martin, M.~Barthelmess, and
  H.~Chapman, \enquote{A scheme for lensless x-ray microscopy combining
  coherent diffraction imaging and differential corner holography,}
  {\protect\JournalTitle{Optics express}} \textbf{20}, 25152--25160 (2012).

\bibitem{bille2016x}
F.~Bill{\`e}, G.~Kourousias, E.~Luchinat, M.~Kiskinova, and A.~Gianoncelli,
  \enquote{X-ray fluorescence microscopy artefacts in elemental maps of
  topologically complex samples: Analytical observations, simulation and a map
  correction method,} {\protect\JournalTitle{Spectrochimica Acta Part B: Atomic
  Spectroscopy}} \textbf{122}, 23--30 (2016).

\bibitem{kourousias2019xrf}
G.~Kourousias, F.~Bill{\`e}, G.~Cautero, J.~Bufon, A.~Rachevski, S.~Schillani,
  D.~Cirrincione, M.~Altissimo, R.~Menk, G.~Zampa, N.~Zampa, I.~Rashevskaya,
  R.~Borghes, M.~Gandola, A.~Picciotto, G.~Borghi, F.~Ficorella, N.~Zorzi,
  P.~Bellutti, G.~Bertuccio, A.~Vacchi, and A.~Gianoncelli, \enquote{Xrf
  topography information: Simulations and data from a novel silicon drift
  detector system,} {\protect\JournalTitle{Nuclear Instruments and Methods in
  Physics Research Section A: Accelerators, Spectrometers, Detectors and
  Associated Equipment}} \textbf{936}, 80--81 (2019).

\bibitem{ippoliti2022reconstruction}
M.~Ippoliti, F.~Bill{\`e}, A.~G. Karydas, A.~Gianoncelli, and G.~Kourousias,
  \enquote{Reconstruction of 3d topographic landscape in soft x-ray
  fluorescence microscopy through an inverse x-ray-tracing approach based on
  multiple detectors,} {\protect\JournalTitle{Scientific Reports}} \textbf{12},
  20145 (2022).

\bibitem{wang2004image}
Z.~Wang, A.~C. Bovik, H.~R. Sheikh, and E.~P. Simoncelli, \enquote{Image
  quality assessment: from error visibility to structural similarity,}
  {\protect\JournalTitle{IEEE transactions on image processing}} \textbf{13},
  600--612 (2004).

\bibitem{bohinc2019nonlinear}
R.~Bohinc, G.~Pamfilidis, J.~Rehault, P.~Radi, C.~Milne, J.~Szlachetko,
  F.~Bencivenga, F.~Capotondi, R.~Cucini, L.~Foglia, C.~CMasciovecchio,
  R.~Mincigrucci, E.~Pedersoli, A.~Simoncig, N.~Mahne, A.~Cannizzo, H.~M. Frey,
  A.~Ollmann, T.~Feurer, A.~A. Maznev, K.~Nelson, and G.~Knopp,
  \enquote{Nonlinear xuv-optical transient grating spectroscopy at the si l2,
  3--edge,} {\protect\JournalTitle{Applied physics letters}} \textbf{114},
  181101 (2019).

\bibitem{yulin2021coatings}
S.~Yulin, M.~Trost, S.~Schwinde, and S.~Schr{\"o}der, \enquote{Coatings with
  barrier layers for extreme-short wavelengths: Euv lithography for the
  semiconductor industry and beyond,} {\protect\JournalTitle{Vakuum in
  Forschung und Praxis}} \textbf{33}, 24--29 (2021).

\bibitem{kujala2019characterizing}
N.~Kujala, M.~Makita, J.~Liu, A.~Zozulya, M.~Sprung, C.~David, and
  J.~Gr{\"u}nert, \enquote{Characterizing transmissive diamond gratings as beam
  splitters for the hard x-ray single-shot spectrometer of the european xfel,}
  {\protect\JournalTitle{Journal of synchrotron radiation}} \textbf{26},
  708--713 (2019).

\bibitem{rosner2018exploiting}
B.~R{\"o}sner, F.~Koch, F.~D{\"o}ring, J.~Bosgra, V.~A. Guzenko, E.~Kirk,
  M.~Meyer, J.~L. Ornelas, R.~H. Fink, S.~Stanescu, S.~Swaraj, R.~Belkhou,
  B.~Watts, J.~Raabe, and C.~David, \enquote{Exploiting atomic layer deposition
  for fabricating sub-10 nm x-ray lenses,}
  {\protect\JournalTitle{Microelectronic Engineering}} \textbf{191}, 91--96
  (2018).

\bibitem{kharitonov2022single}
K.~Kharitonov, M.~Mehrjoo, M.~Ruiz-Lopez, B.~Keitel, S.~Kreis, S.-g. Gang,
  R.~Pan, A.~Marras, J.~Correa, C.~B. Wunderer, and E.~Plönjes,
  \enquote{Single-shot ptychography at a soft x-ray free-electron laser,}
  {\protect\JournalTitle{Scientific Reports}} \textbf{12}, 14430 (2022).

\bibitem{eschen2022material}
W.~Eschen, L.~Loetgering, V.~Schuster, R.~Klas, A.~Kirsche, L.~Berthold,
  M.~Steinert, T.~Pertsch, H.~Gross, M.~Krause, J.~Limpert, and j.~Rothhard,
  \enquote{Material-specific high-resolution table-top extreme ultraviolet
  microscopy,} {\protect\JournalTitle{Light: Science \& Applications}}
  \textbf{11}, 117 (2022).

\end{thebibliography}

\ifarXiv
    

\section*{Supplementary material}

# Three-dimensional coherent diffraction snapshot imaging using extreme ultraviolet radiation from a free electron laser

## Supplementary Material

DANNY FAINOZZI,[1,5,*] MATTEO IPPOLITI,[1,5] FULVIO BILLE',[1] DARIO DE ANGELIS,[1] LAURA FOGLIA,[1] CLAUDIO MASCIOVECCHIO,[1] RICCARDO MINCIGRUCCI,[1] MATTEO PANCALDI,[1] EMANUELE PEDERSOLI,[1] CHRISTIAN M. GÜNTHER,[3] BASTIAN PFAU,[2] MICHAEL SCHNEIDER,[2] CLEMENS VON KORFF SCHMISING,[2] STEFAN EISEBITT,[2,4] GEORGE KOUROUSIAS,[1] FILIPPO BENCIVENGA,[1] AND FLAVIO CAPOTONDI[1]

[1] *Elettra-Sincrotrone Trieste, SS 14 km 163,5 in AREA Science Park, 34149, Trieste, Italy*
[2] *Max-Born-Institut für Nichtlineare Optik und Kurzzeitspektroskopie, Max-Born-Straße 2A, 12489 Berlin, Germany*
[3] *Technische Universität Berlin, Zentraleinrichtung Elektronenmikroskopie (ZELMI), Straße des 17. Juni 135, 10623 Berlin*
[4] *Technische Universität Berlin, Institut für Optik und Atomare Physik, Straße des 17. Juni 135, 10623 Berlin*
[5] *The authors contributed equally to this work.*
*danny.fainozzi@elettra.eu*

## 1. Experimental Setup

Supplementary Figure S1a) sketches the experimental setup: two noncollinear FEL pulses are produced by wavefront beam splitting, by means of the mini-TIMER split, delay and recombination unit, which also allows to precisely control the crossing angle ($\phi$) and relative time delay ($\Delta t$) between the two FEL pulses at the sample [1, 2]. In the present case, $\phi = 40°$ and $\Delta t = 0$, time coincidence was independently verified (with an accuracy better than the FEL pulse duration) by FEL-optical pump-probe measurements, following the same procedure reported in Refs. [1, 2]. Focusing at the sample position was provided by a Kirkpatrik-Baetz adaptive optical system [3] placed upstream of the mini-TIMER system and set in order to have an FEL spot size at the sample of about $60 \times 120$ $\mu m^2$ (horizontal $\times$ vertical) full width at half-maximum (FWHM) for each of the two individual FEL beams. Two CCD detectors (PI-MTE; pixel size $13.5 \times 13.5$ $\mu m^2$, area $2048 \times 2048$ pixels) were placed on-axis with respect to the two FEL pulses at a sample-CCD distance of about 100 mm (see Figure S1a)). Two independent, motorized beamstops, placed in front of each CCD, were used to block the direct FEL beam, even if this was not always necessary in light of the low sample transmission. The FEL output wavelength was tuned to 20.8 nm, typical pulse duration and intensity at the source were about 50 fs (FWHM) and 100 $\mu$J, respectively, while the polarization was circular. During the experiment the FEL intensity at the sample position was reduced by means of solid-state filters along the photon transport to about 1.1 $\mu$J. Due to the different number of reflections in the mini-TIMER system (see Figure S1a), the two FEL pulses had an unbalanced intensity, *i.e.*, about 0.7 $\mu$J for the pulse used to obtain the 40° view and 0.4 $\mu$J for the one crossing the sample at normal incidence. The corresponding average total fluence per shot was of about 15 mJ/cm2, below the sample damage threshold [4].

Figure S1b) shows a scanning electron microscope (SEM) image of one of the 3D structures

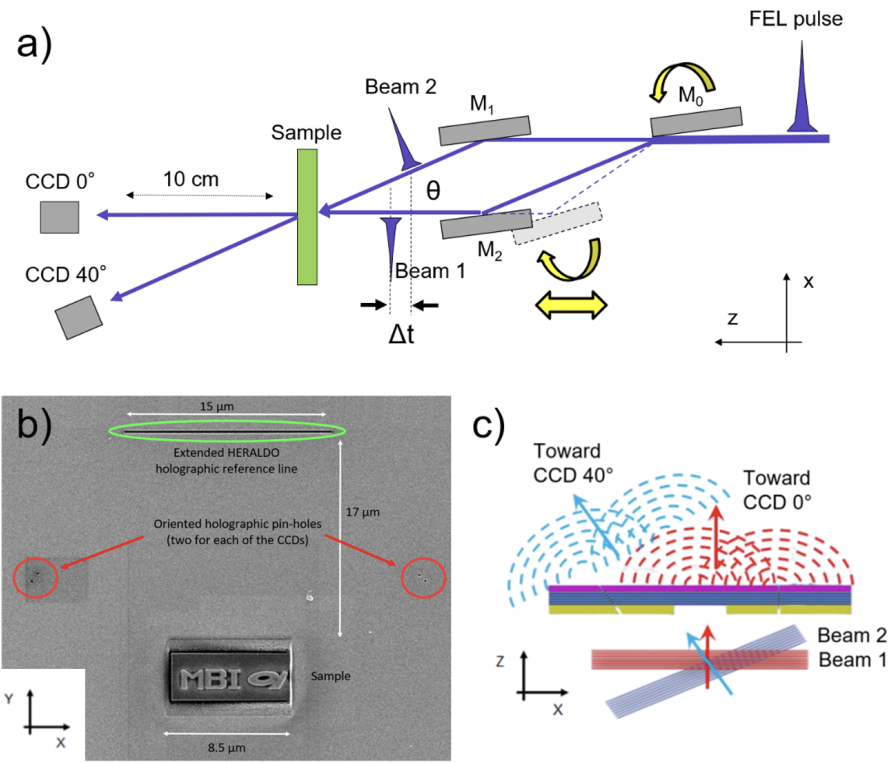

Fig. 1. Panel a) sketches the mini-Timer setup: M0, M1 and M2 are plane mirrors used to split and recombine the FEL beam (blue line) at a given value of $\theta$, the thick yellow arrows and the dashed elements schematize the movements of M0 and M2 used to adjust $\Delta t$ at fixed $\theta$. The reference frame is shown in the bottom-right corner. Panel b) is a picture of one of the employed samples, green ellipse and red circles are, respectively, the slit for HERALDO holography and the pinholes for Fourier transform holography. In the design of the holographic pin-hole reference a set of two pin-holes has been drilled one at normal incidence and one at 40° in order to be parallel to FEL beam 1 (pinhole on the right) and FEL beam 2 (pinhole on the left), as shown in panel c).

used in the experiment. The samples consist of two multi-layers, deposited on the opposite sides of a 30 nm thin silicon nitride membrane, that were subsequently patterned by focused Ga+ ion beam (FIB) milling. One of those layers has a composition of [Cr (5 nm)/ Au (50 nm)]$_5$ and houses the optical mask required for the imaging process as described below. The second layer is a stack of [Al (30 nm)/Ta (2 nm)/ Co$_{0.90}$Tb$_{0.10}$ (20 nm)/ Ta (2 nm)] which serves as the basis out of which the 3D objects were prepared. To achieve the raised 3D structures, the Cr/Au layer was first removed in the intended field of view using a FIB acceleration voltage of 5 kV. Next, approximately 100 nm thick outlines of spiral (see *e.g.* Fig 1 b) and S3 b)) or ramp (see Fig. S1 b)) patterns were cut through the remaining sample stack at 30 kV. These structures can then be raised in a controlled fashion using strain induced by local or global exposure to 30 keV ions. The amount of rise is controlled by the ion dose, where typical values for the structures shown are around 20 pC/$\mu m^2$ [5, 6]. Additional morphological structures, like the "MBI" text label and the "Elettra" logo in Fig. S1b), were produced by electron-beam assisted Pt deposition starting from metalorganic precursors. The approximate thickness of such a carbon-rich Pt layer was about

100 nm. Two pairs of holographic pinholes with a diameter of 90 ± 10 nm were milled along the directions of the two impinging FEL beams at a distance of about 15 $\mu$m from the 3D object, as displayed in Fig. S1b). In such a way, the reference wave generated by each pinhole interferes with only the object-wave for a given beam direction. This allows the two detectors to collect the two independent holograms for the two different side views. Beside the point-like references for holographic reconstructions, an extended line reference 15 $\mu$m long and 100 ± 10 nm wide to perform differential HERALDO holography was milled at a distance of 17 $\mu$m from the 3D structure top border (see Figure S1b) [9,10]. The spatial locations of the holographic reference structures with respect to the central 3D object have been chosen in order to avoid overlapping between the holographic reconstruction and the sample autocorrelation in the inverse Fourier transform process. Inside the experimental chamber, the sample surface was oriented using a five-axis goniometer in order to be orthogonal with respect to the FEL beam labeled as "Beam 1" in Fig. S1a) and S1c). The respective CCD detector (labeled as "CCD 0°") collected a "front view" of the sample, while the second CCD (labeled as "CCD 40°") collected the diffraction pattern at a "tilted view" of 40 degrees.

## 2. CDI Reconstruction

For each pair of diffraction patterns simultaneously collected, we subtracted the respective background images, acquired with the same integration time but with the FEL beam off. These images contain the information on the dark count of detectors and residual background light in the experimental chamber. The Fourier transform of the diffraction pattern provides the cross-correlations between the reference pinholes and the sample (holographic approach), while the HERALDO approach essentially consists in introducing a numerical differential filter along the direction of the extended reference before the Fourier transform [7, 8]. Holographic inversion of raw data successfully provides low-resolution images of the sample, which are sufficiently clear to demonstrate how independent visions of the same object from different sides can be obtained. This also indicates that cross-contamination of the two diffraction patterns is minimal. In order to improve the image quality, two isomorphic transformations are used to correct for both the artificial intensity modulation modulation (see Fig. S2 c)) and the possible angular misalignment between the HERALDO reference and the differential filter operator used to retrieve the spatial image (see Fig. S2 e)). The first transformation corrects for the stereographic projection of the diffraction pattern, defined by the Ewald sphere, onto the flat plane of the CCD chip [9, 10]. The larger is $\phi$ and the larger is this correction. In the case of the 40° view, this correction considerably stretches the image in both directions, resulting into a loss of signal at the highest $|q|$'s. This is partially prevented by numerically zero padding the equivalent of 200 pixels to the diffraction image before applying this correction. The second transformation of the diffraction pattern is a rotation around the maximum of the diffraction peak, which accounts for the aforementioned positional misalignment, which, in turn, may result in aberrations of the reconstructed image when not properly corrected. Indeed, we experimentally observe (Fig. S2 c)) that the sample was azimuthal rotated by ∼ 5° with respect to the x axis (the reference frame is reported in Fig. S1). The beneficial effects of these corrections are shown in Fig. S2, where starting from a raw diffraction pattern (Fig. S2a)), the correction for the Ewald sphere projection and sample rotation are shown for the 40° view (Fig. S2b)). The improved quality both in terms of intensity uniformity inside the voids of the lobes, as well as the reduction of the stripes in the HERALDO reconstruction, can be clearly noticed in Fig.s S2d) – S2f), where most of the raw data artifacts (see Fig. S2c) and S2d)) are efficiently corrected. Figure S3 sketches the CDI phasing loop process. The holograms reconstructed as described above are used as starting guesses of the object and to define the binary support mask (Figs. S3a) and S3b)). The phase retrieval algorithm, schematically represented in Fig. S3c), first performs the Fourier transform of the nth iteration object, then substitutes the amplitude of the Fourier transform with the square root of the measured

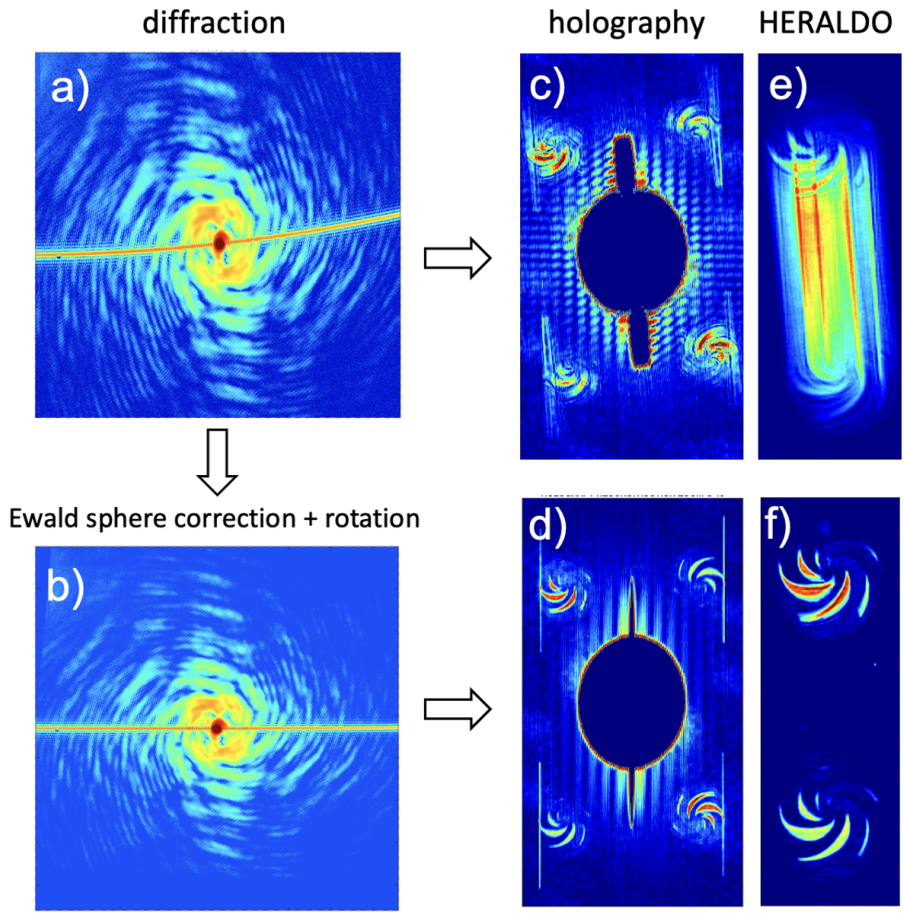

Fig. 2. Panel a) shows the diffraction pattern collected at 40° tilted view (see Figures S1a and S1c) from the five-fold helicoidal structure shown in Fig. 2b) in the MT. In order to remove undesired aberrations, a correction by the Ewald sphere must be performed before the phase retrieval reconstruction of the sample. On our data, a further correction for the tilt of the sample has been implemented, as shown in panel b)). Panels c), d), e) and f) show a comparison of the holographic and HERALDO inversions with and without correction.

diffraction pattern. After this iteration, an inverse Fourier transform is applied to return to the real space. Finally, the $(n + 1)^{th}$ object is upgraded, according to the employed algorithm, using the binary support mask. In the CDI phase retrieval algorithm, we use a combination of the Error Reduction (ER) algorithm, the Hybrid Input-Output (HIO) algorithm, and the Relaxed Averaged Alternating Reflections (RAAR) algorithm [11, 12]. The first 10 iterations were performed using the ER algorithm, then a sequence of 10 iterations for ER, HIO and RAAR were repeated 30 times. Finally, the last 150 iterations were performed using the ER algorithm only for a total of 1060 iterations. This number was set as the lowest number of iterations for which the reconstruction algorithm reaches convergence. During the sequence of ER, HIO and RAAR the support was updated 5 times using a Gaussian filter on the reconstructed object. Our algorithm was written in MATLAB (version R2021a (9.10.0)), an iteration took at most 2 seconds, and all iterations took less than 35 minutes for each view.

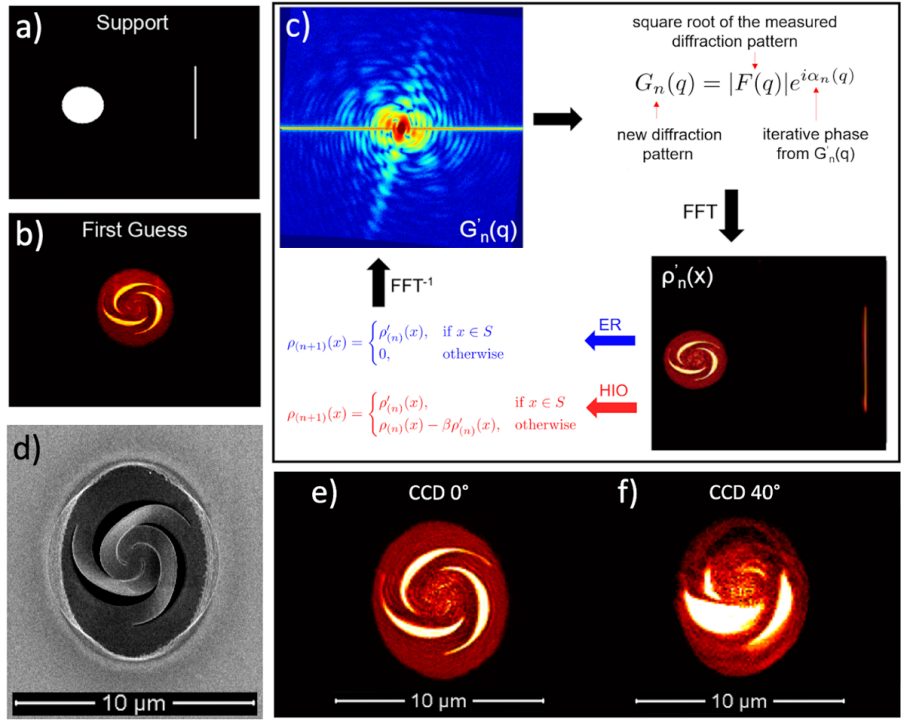

Fig. 3. Panel c) shows a sketch of the reconstruction algorithm; the support S and the first guess are used to iteratively reconstruct the diffraction pattern $G(q)$ of the sample. At the Nth iteration the amplitude of the diffraction pattern is taken as the measured spectrum while the phase is changed. Different constraints are applied to the real space reconstruction function $\rho(x)$ depending on the kind of algorithm chosen. Panel d) and e) show the reconstructed images of the sample for the two CCDs for the 0° and 40° directions, respectively.

## 3. 3D reconstruction: transmission and phase evaluation

The relative transmission image $I/I_0$ that is obtained from the CDI reconstruction can in principle be calculated using the Beer-Lambert's law of absorption [13]:

$$I = I_0 e^{-\mu(\lambda)\rho l} \quad (1)$$

where $I_0$ is the incident beam's intensity, I is the intensity after the sample, $\mu$ is the attenuation coefficient at a given wavelength $\lambda$, $\rho$ is the sample density and $l$ is the length of the path traveled by the photons in the sample. In Eq. 1, both $\mu$ and $\rho$ depend on the longitudinal coordinate along the direction of the incidence radiation and consider the different materials crossed by EUV radiation. Eq. 1 is at the core of the simulation of the $I/I_0$ images, which is used in the 3D sample evaluation process to establish the fidelity of the reconstructed object, by confronting the simulated transmission with the actual $I/I_0$ map from the experiment. Furthermore, since we have a detailed knowledge of the sample composition reported in Section 1, the photon path l can be uniquely retrieved. As a result, initial thickness maps associated with the two views can be calculated to start the sample reconstruction process. For some of the investigated samples, the thickness is large enough to practically suppress the transmission. When this happens, the algorithm sets a value of l such as $I/I_0 = 0.001$ in Eq. 2. This value was chosen so that the computed maximum thickness would not exceed the expected thickness of the samples. The

phase images $\phi$ are instead simulated, starting from a 3D sample structure, using Eq. 2 [14]:

$$\phi = \lambda r_e \int_{l_p} dl \rho_e(l) \qquad (2)$$

where $r_e = 2.818 \times 10^{-15}$ m is the classical radius of the electron and $\rho_e$ is the electronic density (for the given specimen $\rho_e = 1.3 \times 10^{30}$ m$^{-3}$). The capability of simulating phase responses is an additional constraint, beside the view dependent $I/I_0$ transmission maps, in the evaluation of the reconstructed 3D sample, thereby identifying the most likely solution.

## 4. 3D reconstruction: choosing the optimal solution

The initial step in the 3D reconstruction process involves the calculation of thickness maps from two different views using Eq. 2. These maps are then combined together to generate multiple 3D test models considering both transmission and phase information through the application of a recently developed ray tracing method [15]. Each resulting 3D model must meet the requirement of continuity along view directions, meaning that there can be no voids or gaps inside the sample along the imaging directions. The absolute difference between the thickness of the generated models and the sample thickness maps is evaluated for both views, and if the average absolute difference exceeds a predetermined threshold (in this case, 20%), the solution is rejected. After this filter process based on qualitative considerations, in order to determine the accuracy and fidelity of the other generated 3D models, a forward simulation algorithm, that use Eq. 1 and 2 to reproduce the CDI reconstructed images of the transmission ($I/I_0$) and phase ($\phi$) maps, for the two independent views, is adopted. The four synthetic images are then evaluated in order to assign a score for each possible 3D sample solution and establish which one is the best approximation of the real underlying structure. This is accomplished through the use of the structural similarity index measure (SSIM) [16], which is a perceptual metric that quantifies the image quality, associating a degradation rank, between a target and a reference image. Once all SSIM computations have been carried out for $I/I_0$ and $\phi$ images, their scores are summed together and the test sample producing the largest sum is chosen as the likely solution of 3D reconstructed object. The presented algorithm was developed in-house with Python (Python Software Foundation). A limitation in the implementation of the 3D reconstruction algorithm lies in its computational demand. Although the code has been designed to parallelize the generation and evaluation of all test model samples, and the core functions have been compiled through the Numba just-in-time compiler [17], the algorithm can still be computationally demanding, depending on the size of the input views. For example, to find the solution of the 3D object in Fig.s 3 of the main text the average computational time for each sample was 137 min on a workstation with 8 threads CPU (i7-6700 at 3.40 GHz) and 48 GB of RAM. In order to demonstrate the robustness of the 3D reconstruction algorithm, independently to the spatial frequency content of the sample, in addition to the structure reported in the main text, Figs. S4a) and S4b) show respectively the 2D and 3D reconstructions of the sample displayed in Fig. S1b). The CDI views have been obtained averaging 30 different datasets acquired exposing the object to multiple FEL shots, for a total of 750 exposures. Additional 3D reconstructed objects are shown in Fig. S4c) and S4d), where we report the results obtained from the five-fold helicoidal structure of the MT and a three-fold helicoidal structure visible in Figs S3d) and S4d). In all of these three representative cases the combination of holographically-guided CDI and 3D ray tracing projection algorithms is able to reconstruct a realistic solution of the illuminated object. As shown in more detail in section 5, the image quality will not degrade by reducing the number of images or by reducing the number of FEL shots for each image, showing the computational robustness of the developed stereographic retrieval procedure. As an example, Figs. S5a) and S5b) show the counterpart of Figs S4a) and S4b) for a single image corresponding to 25 FEL shots. The scale bars displayed in the 3D reconstructions of Figs. S4 and S5 were constructed

using $L_{depth} = 33 \pm 5$ nm and aid in understanding how the level of details are comparable with the theoretically estimated CDI depth resolution of about 90 nm.

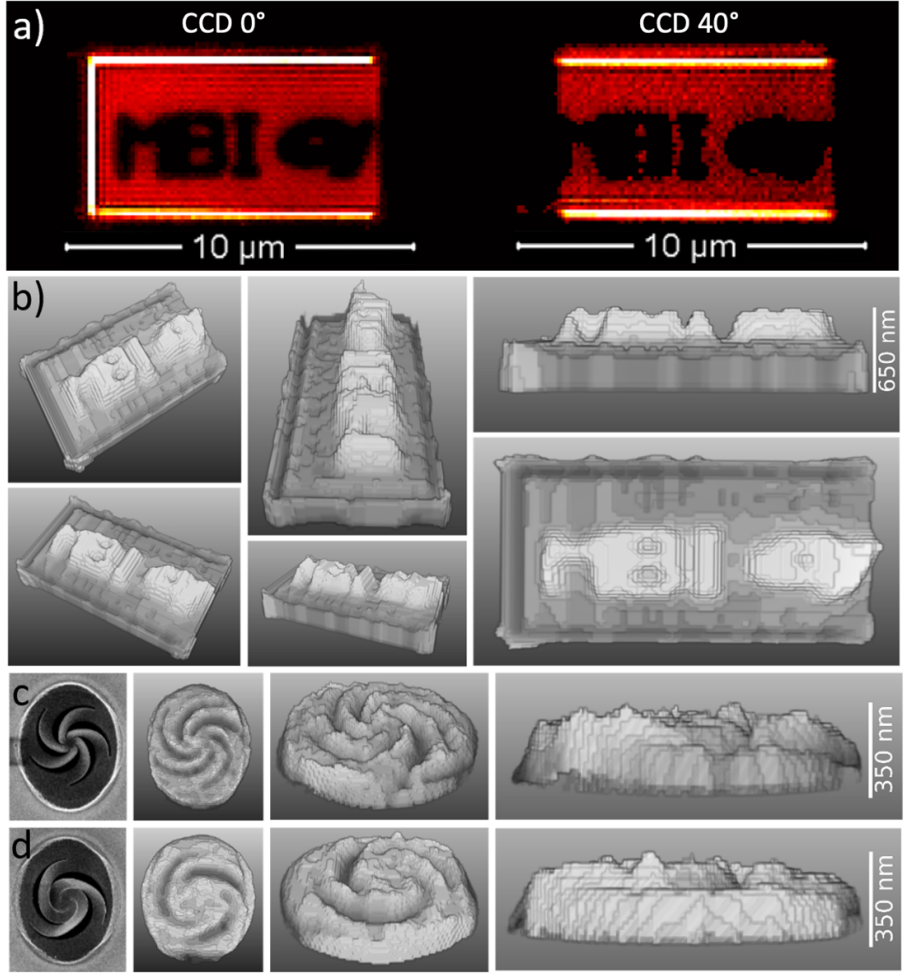

Fig. 4. a) CDI 2D reconstructions obtained by averaging 30 datasets each one recorded by exposing the sample to 25 FEL pulses. b) corresponding 3D reconstruction obtained with ray tracing code. c) and d) reconstructions of the two helicoidal samples shown in the SEM images on the left.

## 5. Phase retrieval transfer function

Under the experimental scattering geometry described in Section 1, the maximum value of the exchanged momentum $|q| = (2/\lambda) \sin(\alpha/2)$, where $\lambda$ is the FEL wavelength and $\alpha$ the CCD collection angle, is $|q|_{max} = 6.6$ $\mu m^{-1}$ at the midpoint of the detector edge, corresponding to a theoretical diffraction-limited lateral resolution $L_{res} = \lambda/(2 \cdot \sin(\alpha_{max})) = 76$ nm. Such theoretical lateral resolution can be improved if the maximum collection angle in the CCD detector is considered along the diagonal direction of the sensor chip, in such case, the increased acceptance angle gives $|q|_{max} = 9.3$ $\mu m^{-1}$ corresponding to a diffraction limited spatial resolution of $L_{res} = 54$ nm. Due to the detector noise, the stray radiation from the beamline and the not perfect transversal coherence of the beam, the real resolution of CDI reconstruction may be worse

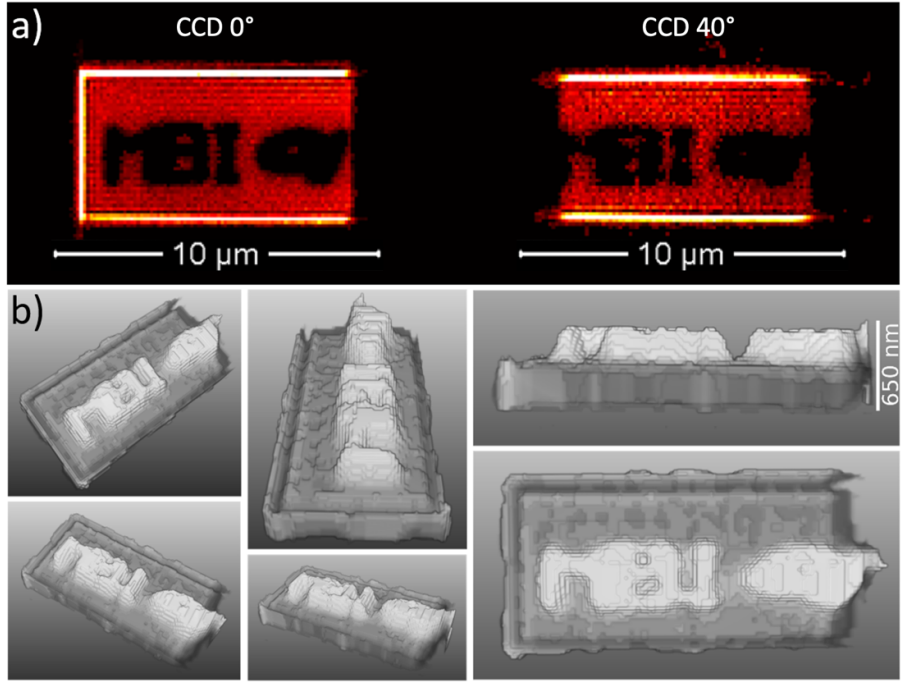

Fig. 5. a) CDI 2D reconstructions obtained using a single image obtained by exposing the sample to 25 FEL pulses. b) corresponding 3D reconstruction obtained with ray tracing code.

than $L_{\text{res}}$, determined by geometric considerations. Indeed, if the retrieved phases are incorrect, the retrieved image quality might degrade, in a similar way as phase errors in a lens can cause aberrations in optical microscopy. In order to account for that, following Ref. [18] we estimate the CDI resolution by using the phase retrieval transfer function (PRTF), which evaluates the combined effect of the signal-to-noise ratio (SNR) of the diffracted wave and the systematic errors that corrupt the convergence of the retrieval algorithm, and is defined as:

$$\text{PRTF} = \frac{\left\langle \left| \left\langle E_{ij}^{\text{rec}}(k_x, k_y) \right\rangle_j \right| \right\rangle}{\left\langle \left| E_i^{\text{meas}}(k_x, k_y) \right| \right\rangle} \qquad (3)$$

where $E^{\text{meas}}$ is the magnitude of the measured electric field at the detector, corresponding to spatial frequency $(k_x, k_y)$, $E^{\text{rec}}$ is the reconstructed electric field at the $j^{\text{th}}$ iteration, corresponding to spatial frequency $(k_x, k_y)$, while the index $i = (1, 2)$ labels the two stereoscopic views. The PRTF thus compares the Fourier amplitudes of the average of multiple subsequent reconstructions with the measured diffraction amplitudes, and represents the confidence with which the diffraction phase is stably retrieved without ambiguity among successive iterations of CDI code. Indeed, where the phase of a particular Fourier component is consistently retrieved, the complex values add in phase and hence the PRTF is equal to unity, whereas if the phase is random the sum approaches zero and the PRTF vanishes. In addition to analysis reported in Fig.4a) of the main text, Fig S6 shows the comparison of the PRTF averaged over shells of constant values of $|q|$ for the reconstructed sample reported in Figs. S1b). Due to a higher symmetry of the lamella sample in Fig S1a), the diffraction pattern is less sparse on CCD chip with respect to three and five-fold helicoidal structures, and as a consequence the PRTF function decrease faster as a

function of $|q|$. In such a case, according to [18], we consider the typical $1/e$ threshold to define the lateral resolution for the PRTF analysis. The PRTF function indicates that the resolution slightly decreases while reducing the number of accumulated shots, from about 110 nm for 750 FEL exposures to about 180 nm for 25 FEL shots. The PRTF evaluation of the image lateral resolution is consistent with the ability of CDI to resolve the fine features inside the "MBI" text, as displayed in Figs S4a) and S5a). Indeed, comparing the SEM image in Fig. S1b) with the CDI images, the letters "B" and "I" are well spatially resolved while no discontinuity is observed between the "M" and "B" in the phase retrieved objects. From SEM images we estimate that the minimum distance between "B" and "I" is about 200 nm while "M" and "B" are separated by 100 nm, in quite fair agreement with the PRTF result.

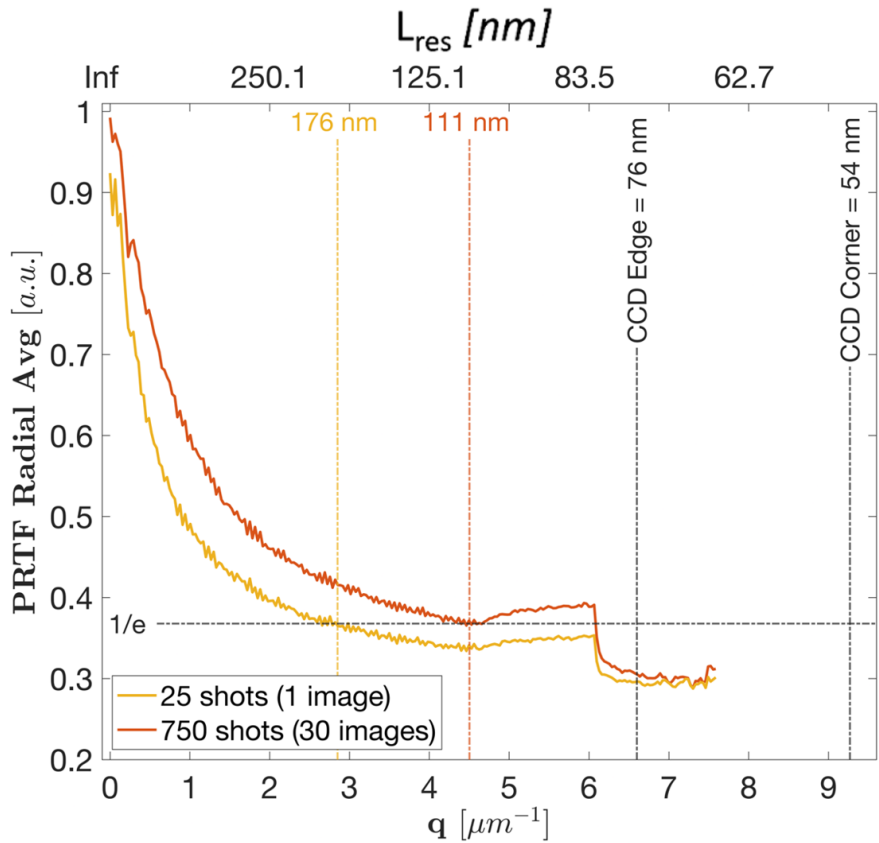

Fig. 6. PRTF radial average functions for the "MBI+Elettra" sample as a function of FEL illumination shots. Yellow and Red curves show the PRTF function for a single 25-shot exposure, and the average of 30 different 25-shot exposures, respectively. The lateral resolution is found to be in the range of about 100 nm for the averaged images to about 180 nm for the single image, according to the empiric PRTF = $1/e$ level.

## References

1. F. Bencivenga, R. Cucini, F. Capotondi, A. Battistoni, R. Mincigrucci, E. Giangrisostomi, A. Gessini, M. Manfredda, I. Nikolov, E. Pedersoli, E. Principi, C. Svetina, P. Parisse, F. Casolari, M. B. Danailov, M. Kiskinova, and C. Masciovecchio, "Four-wave mixing experiments with extreme ultraviolet transient gratings," Nature **520**, 205–208 (2015).

2. R. Mincigrucci, L. Foglia, D. Naumenko, E. Pedersoli, A. Simoncig, R. Cucini, A. Gessini, M. Kiskinova, G. Kurdi, N. Mahne, M. Manfredda, I. P. Nikolov, E. Principi, L. Raimondi, M. Zangrando, C. Masciovecchio, F. Capotondi, and F. Bencivenga, "Advances in instrumentation for fel-based four-wave-mixing experiments," Nucl. Instruments Methods Phys. Res. Sect. A: Accel. Spectrometers, Detect. Assoc. Equip. **907**, 132–148 (2018).
3. L. Raimondi, M. Manfredda, N. Mahne, D. Cocco, F. Capotondi, E. Pedersoli, M. Kiskinova, and M. Zangrando, "Kirkpatrick–baez active optics system at fermi: system performance analysis," J. synchrotron radiation **26**, 1462–1472 (2019).
4. T. Wang, D. Zhu, B. Wu, C. Graves, S. Schaffert, T. Rander, L. Müller, B. Vodungbo, C. Baumier, D. P. Bernstein, B. Bräuer, V. Cros, B. De Jong, R. Delaunay, A. Fognini, R. Kukreja, S. Lee, V. López-Flores, J. Mohanty, B. Pfau, H. Popescu, M. Sacchi, A. B. Sardinha, F. Sirotti, P. Zeitoun, M. Messerschmidt, J. J. Turner, W. F. Schlotter, O. Hellwig, R. Mattana, N. Jaouen, F. Fortuna, Y. Acremann, C. Gutt, H. A. Dürr, E. Beaurepaire, C. Boeglin, S. Eisebitt, G. Grübel, J. Lüning, J. Stöhr, and A. O. Scherz, "Femtosecond single-shot imaging of nanoscale ferromagnetic order in co/pd multilayers using resonant x-ray holography," Phys. review letters **108**, 267403 (2012).
5. J. Li and Z. Liu, "Focused-ion-beam-based nano-kirigami: from art to photonics," Nanophotonics **7**, 1637–1650 (2018).
6. S. Chen, J. Chen, X. Zhang, Z.-Y. Li, and J. Li, "Kirigami/origami: unfolding the new regime of advanced 3d microfabrication/nanofabrication with "folding"," Light. Sci. & Appl. **9**, 75 (2020).
7. M. Guizar-Sicairos and J. R. Fienup, "Holography with extended reference by autocorrelation linear differential operation," Opt. express **15**, 17592–17612 (2007).
8. M. Guizar-Sicairos and J. R. Fienup, "Direct image reconstruction from a fourier intensity pattern using heraldo," Opt. letters **33**, 2668–2670 (2008).
9. R. Sandberg, D. Raymondson, A. Paul, K. Raines, J. Miao, M. Murnane, H. Kapteyn, and W. Schlotter, "Tabletop soft-x-ray fourier transform holography with 50 nm resolution," Opt. letters **34**, 1618–1620 (2009).
10. S. Schaffert, B. Pfau, J. Geilhufe, C. M. Günther, M. Schneider, C. von Korff Schmising, and S. Eisebitt, "High-resolution magnetic-domain imaging by fourier transform holography at 21 nm wavelength," New J. Phys. **15**, 093042 (2013).
11. J. R. Fienup, "Phase retrieval algorithms: a comparison," Appl. optics **21**, 2758–2769 (1982).
12. D. R. Luke, "Relaxed averaged alternating reflections for diffraction imaging," Inverse problems **21**, 37 (2004).
13. D. F. Swinehart, "The beer-lambert law," J. chemical education **39**, 333 (1962).
14. G.-H. Chen, J. Zambelli, N. Bevins, Z. Qi, and K. Li, "X-ray phase sensitive imaging methods: basic physical principles and potential medical applications," Curr. Med. Imaging **6**, 90–99 (2010).
15. M. Ippoliti, F. Billè, A. G. Karydas, A. Gianoncelli, and G. Kourousias, "Reconstruction of 3d topographic landscape in soft x-ray fluorescence microscopy through an inverse x-ray-tracing approach based on multiple detectors," Sci. Reports **12**, 20145 (2022).
16. Z. Wang, A. C. Bovik, H. R. Sheikh, and E. P. Simoncelli, "Image quality assessment: from error visibility to structural similarity," IEEE transactions on image processing **13**, 600–612 (2004).
17. S. K. Lam, A. Pitrou, and S. Seibert, "Numba: A llvm-based python jit compiler," in *Proceedings of the Second Workshop on the LLVM Compiler Infrastructure in HPC,* (2015), pp. 1–6.
18. H. N. Chapman, A. Barty, M. J. Bogan, S. Boutet, M. Frank, S. P. Hau-Riege, S. Marchesini, B. W. Woods, S. Bajt, W. H. Benner, R. A. London, E. Plönjes, M. Kuhlmann, R. Treusch, S. Düsterer, T. Tschentscher, J. R. Schneider, E. Spiller, T. Möller, C. Bostedt, T. Hoener, D. A. Shapiro, K. O. Hodgson, D. Van Der Spoel, F. Burmeister, M. Bergh, C. Caleman, G. Huldt, M. M. Seibert, F. R. N. C. Maia, R. W. Lee, A. Szöke, N. Timneanu, and J. Hajdu, "Femtosecond diffractive imaging with a soft-x-ray free-electron laser," Nat. Phys. **2**, 839–843 (2006).


\fi






\end{document}